\documentclass[sigconf,nonacm]{acmart}

\AtBeginDocument{%
  }

\copyrightyear{2022}
\acmYear{2022}
\setcopyright{acmcopyright}\acmConference[RecSys '22]{Sixteenth ACM Conference on Recommender Systems}{September 18--23, 2022}{Seattle, WA, USA}
\acmBooktitle{Sixteenth ACM Conference on Recommender Systems (RecSys '22), September 18--23, 2022, Seattle, WA, USA}
\acmPrice{15.00}
\acmDOI{10.1145/3523227.3546775}
\acmISBN{978-1-4503-9278-5/22/09}

\acmJournal{JACM}
\acmVolume{37}
\acmNumber{4}
\acmArticle{111}
\acmMonth{8}

\usepackage{caption}
\usepackage{subcaption}
\usepackage{multirow}

\begin{document}

\title{Learning Recommendations from User Actions in the Item-poor Insurance Domain}

\author{Simone Borg Bruun}
\email{simoneborgbruun@di.ku.dk}
\orcid{0000-0003-1619-4076}
\affiliation{%
  \institution{University of Copenhagen}
  \city{Copenhagen}
  \country{Denmark}
}

\author{Maria Maistro}
\email{mm@di.ku.dk}
\orcid{0000-0002-7001-4817}
\affiliation{%
  \institution{University of Copenhagen}
  \city{Copenhagen}
  \country{Denmark}
}

\author{Christina Lioma}
\email{c.lioma@di.ku.dk}
\orcid{0000-0003-2600-2701}
\affiliation{%
  \institution{University of Copenhagen}
  \city{Copenhagen}
  \country{Denmark}
}

\authorsaddresses{}

\renewcommand{\shortauthors}{Bruun et al.}

\begin{abstract}
 While personalised recommendations are successful in domains like retail, where large volumes of user feedback on items are available, the generation of automatic recommendations in data-sparse domains, like insurance purchasing, is an open problem. 
The insurance domain is notoriously data-sparse because the number of products is typically low (compared to retail) and they are usually purchased to last for a long time. Also, many users still prefer the telephone over the web for purchasing products, reducing the amount of web-logged user interactions. 
To address this, we present a recurrent neural network recommendation model that uses past user sessions as signals for learning recommendations. Learning from past user sessions allows dealing with the data scarcity of the insurance domain. Specifically, our model learns from several types of user actions that are not always associated with items, and unlike all prior session-based recommendation models, it models relationships between input sessions and a target action (purchasing insurance) that does not take place within the input sessions. Evaluation on a real-world dataset from the insurance domain (ca. $44$K users, $16$ items, $54$K purchases, and $117$K sessions) against several state-of-the-art baselines shows that our model outperforms the baselines notably. Ablation analysis shows that this is mainly due to the learning of dependencies across sessions in our model. We contribute the first ever session-based model for insurance recommendation, and make available our dataset to the research community.
\end{abstract}



\maketitle

\section{Introduction}

Within the domain dealing with insurances for individuals such as home insurance, car insurance and accident insurance, personalised recommendations can help customers continuously adjust their insurances to suit their needs.

Recommender systems (RSs) most often learn from user feedback on items such as books purchased, movies watched, and/or ratings given to those items. These systems need a considerable amount of previous feedback on items to give high quality recommendations. This is a problem in
the insurance domain, where typically there are few different items to purchase, and insurance products typically are bought to be used for a long time, so the purchase frequency is low.

Another problem in insurance recommendation is that several state-of-the-art RSs, e.g., collaborative filtering and content-based, assume that user preferences are static. However, user needs in insurance evolve over time as they are closely connected to life events like marriage, birth of child, buying a home, changing jobs, retirement, etc.

To deal with the above idiosyncrasies of the insurance domain, we focus on a session-based approach to model insurance recommendations. Session-based RSs do not use information about long-term user preferences. Instead recommendations are based on short-term user-item interactions within the ongoing session. In that way, session-based RSs capture dynamic user preferences using the session as temporal context for the recommendation. However, current session-based RSs cannot be applied to our insurance domain because many users buy insurance products over the phone \textit{after having sessions} on the insurance website. Thus, when using past user sessions on the insurance website as signals for learning recommendations, the target action (i.e., the user's purchase of items) does not happen within the input sessions. This is an important difference from the usual session-based task, where the target action is the next item the user interacts with \textit{in the ongoing session}.

Session-based RSs usually learn from user interactions with items e.g., view of videos or books added to cart. However, on an insurance website there are several other user actions that can be useful signals for learning recommendations. For instance, if a user has recently tried to report a claim that the user’s current insurance products did not cover or if a user has recently updated his/her employment at the personal account. We thus explore the usefulness of such signals by learning recommendations from several types of user actions that are not always directly associated with items.

We present a recurrent neural network (RNN) recommendation model that learns from several types of user actions, and unlike all prior session-based recommendation models, it models relationships between input sessions and a target action that does not take place within the input sessions. 
We call this approach a \emph{cross-sessions} recommendation model.
Evaluation on a real-world dataset from the insurance domain 
and against several state-of-the-art baselines shows that our model outperforms the baselines notably. 
We \textbf{contribute} the first ever cross-sessions model for insurance recommendation and make our dataset publicly available.

\section{Related Work}

We present related work focusing on RSs within the insurance domain (Section~\ref{subsec:related_insurance}) and session-based RSs (Section~\ref{subsec:related_session}).

\subsection{Insurance Domain}
\label{subsec:related_insurance}

Previous research on RSs for the insurance domain is limited. In principle, a knowledge-based RS could be used for this task, which would mine highly personalised user information through user interactions \cite{HuiZZWN22}, however to our knowledge no prior work reports this. Most of it supplements the small volume of user feedback with user demographics, such as age, marital status, income level, and many more. We overview these next.

\citet{Xu2014} cluster users into different groups based on their demographics. Then they make association rule analysis within each group on the users’ set of purchased items. Recommendations are extracted directly from the association rules. \citet{Mitra2014} estimate similarities between users based on demographic attributes using a similarity measure (e.g., cosine similarity). Then they make recommendations to a user based on the feedback on items by the top-N similar users. \citet{Qazi2017, Qazi2020} train a Bayesian network with user demographics and previously purchased items as input features, aiming to predict the last purchased item of a user. They further train a feed forward neural network to provide recommendations to potential users where only external marketing data is available. All these methods show more effective recommendations compared to standard RS approaches, such as matrix factorization and association rule mining solely applied to the feedback data. In addition, these methods solve the problem of cold start users occurring when recommending items to users with no previous feedback on items. Unlike these methods above, which assume that the preference dynamics are homogeneous within demographic segments, our model allows the changes in user preferences to be individual by using sessions generated by the individual user.

\citet{Bi2020} propose a cross-domain RS for the insurance domain. They use knowledge from an e-commerce source domain with daily necessities (clothes, skincare products, fruits, electronics products, etc.) to learn better recommendations in the insurance target domain when data is sparse. They employ a Gated Recurrent Unit (GRU)~\cite{ChoEtAl2014} to model sequential dependencies in the source domain. Our model differs from this model by using user sessions on the website in the target domain, thereby not having the need of overlapping users across multiple domains. Moreover, the model in \cite{Bi2020} is not session-based, but it is based on users' long-term preferences in both source and target domain.

\subsection{Session-based Recommender Systems}
\label{subsec:related_session}

Session-based RSs model users' short-term preferences within the ongoing session~\cite{FangEtAl2020}, commonly using item-to-item recommendations~\cite{Davidson2010,Linden2003}: similarities between items are computed based on the session data, i.e., items that are often interacted with together in sessions achieve a high similarity. The item similarities are then used during the session to recommend the most similar items to the item that the user has currently interacted with. This approach only considers the last interaction of the user, ignoring even the information of the past interactions in the ongoing session. Moreover, this approach requires the target action to happen within sessions, thus this approach cannot be applied to our task.

An extension of the item-to-item approach is Session-based K-Nearest Neighbors (SKNN)~\cite{JannachAndLudewig2017,HuEtAl2020}, which considers all user interactions in the session. In this approach similarities between entire sessions are computed using a similarity measure (e.g., cosine similarity). The recommendations are then based on selecting items that appeared in the most similar past session. This approach does not take into account the order of the input sequence.

 A sequential extension to the SKNN method is Vector Multiplication SKNN~\cite{LudewigAndJannach2018}, which puts more weight on the most recent user interactions of a session when computing the similarities. Another extension is Sequence and Time Aware Neighborhood~\cite{GargEtAL2019}, which uses the position of an item in the current session, the recency of a past session with respect to the current session, and the position of a recommendable item in a neighbouring session. The latter model depends on the target actions occurring within sessions, and so does not fit our task where the target actions occur outside the session.

Neural session-based approaches have also been proposed. One of the most cited is GRU4REC~\cite{HidasiEtAl2016,HidasiAndKaratzoglou2018}. This approach models user sessions with the help of GRU in order to predict the probability of the subsequent interaction given the sequence of previous interactions. Neural Attentive Session-based Recommendation~\cite{LiEtAl2017} extends GRU4REC by using an attention mechanism to capture the user’s main purpose in the current session. Moreover, Graph Neural Networks have been used ~\cite{WuEtAl2019}. This approach models session sequences as graph structured data, thereby being capable of capturing transitions of items and generating item embedding vectors correspondingly.

The above methods use only the user's single ongoing session. Attempts to use past user sessions when predicting the next interaction for the current session have also been proposed~\cite{Quadrana2017, Ruocco2017, Ying2018, Hu2018, Phuong2019}. However, an in-depth empirical investigation of these methods~\cite{Latifi2021} shows that they did not improve over heuristic extensions of existing session-based algorithms that e.g., extend the current session with previous sessions or boost the scores of items previously interacted with. Unlike all above methods, we use different types of actions, not only with items (see Section~\ref{subsec:dataset}).

\section{Approach}
We present the problem formulation (Section~\ref{subsec:problem}) and our model for addressing it (Section~\ref{subsec:approach}).

\subsection{Problem Formalization}
\label{subsec:problem}

\begin{figure}[tb]
    \centering
    \includegraphics[width=0.48\textwidth]{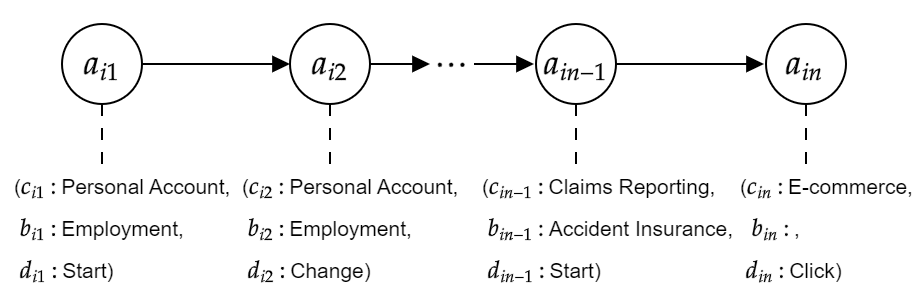}
    \caption{Example of session on insurance website. A session is a list of 3-tuple actions in the order of time.}
    \label{fig:insurance session}
\end{figure}

The goal of our cross-sessions RS is to recommend the next items that a user will buy, given the user's past sessions. 
As opposed to standard session-based RSs, which predict the next step in the input sequence given the sequence so far, in our task: (1) the target action, i.e., purchase, occurs outside the session; (2) we use user actions across multiple sessions; (3) we use all user actions, not only actions with items.
We extend the notation in~\cite{FangEtAl2020} to accommodate these differences.

A session, $s_i$, is a sequence of user actions, $\lbrace a_{i1},a_{i2},a_{i3},...,a_{in} \rbrace$, on the website. An action, $a_{ij}$, is represented by a 3-tuple, i.e., $a_{ij} = (c_{ij}, b_{ij}, d_{ij})$, where:
\begin{itemize}
    \item $c_{ij}$: action section, refers to the section of the website in which the user interacts;
    \item $b_{ij}$: action object, refers to the object on the website that a user chooses to interact with; and
    \item $d_{ij}$: action type, refers to the way that a user interacts with objects.
\end{itemize}
Fig.~\ref{fig:insurance session} illustrates a session where the user starts (action type $d_{i1}$) by interacting with employment (object $b_{i1}$) in the personal account section of the website (section $c_{i1}$).
Then, the user changes ($d_{i2}$) the employment ($b_{i2}$) in the same personal account section ($c_{i1}$).
Section~\ref{subsec:dataset} and Tab.~\ref{tab:actions} present different sections, objects, and action types. 

The past sessions of a user is a list of sessions, $s_i$, chronologically ordered. We do not include all historical sessions of a user as we assume that only recent sessions are relevant for the current task. We use an inactivity threshold, $t$, to define recent sessions, and reason that two sessions belong to the same task if there is no longer than $t$ (time) between them.
We describe how we estimate the threshold $t$ from real data in Section~\ref{subsec:dataset}. The task is to learn a function, $f$, for predicting the probability that a user will buy each item $k$ after the last session $s_m$ based on the input sequence of user's past sessions:
\begin{equation}
    f(s_1,s_2,s_3,...,s_m) = (\hat{p}_1,\hat{p}_2,\hat{p}_3,...,\hat{p}_K),
    \label{eq:task}
\end{equation}
where each element in $\lbrace s_1,s_2,s_3,...,s_m \rbrace$ is a sequence of actions, $\hat{p}_k$ is the estimated probability that item $k$ will be bought by the user, and $K$ is the total number of items.

\subsection{Proposed Approach}
\label{subsec:approach}

\begin{figure*}[tb]
\centering
\begin{minipage}[t]{.22\textwidth}
    \includegraphics[width=\linewidth]{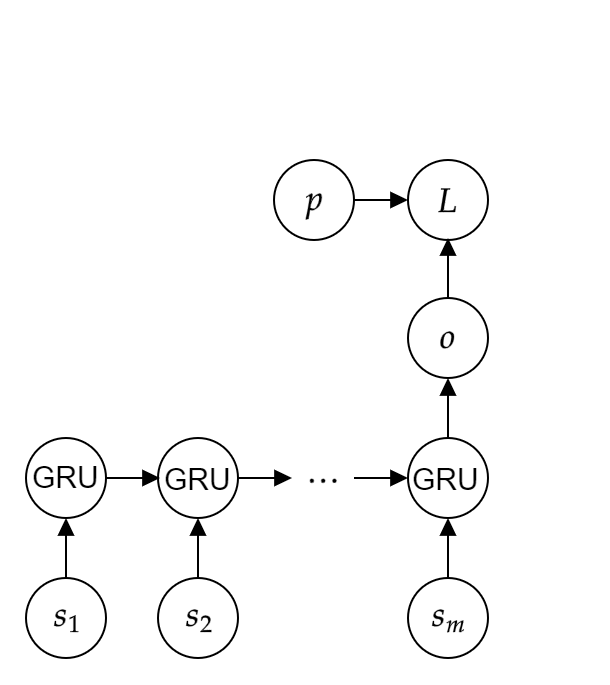}
    \caption{Architecture of cross-sessions encode}
    \label{fig:architecture_encode}
\end{minipage}%
\begin{minipage}[t]{.44\textwidth}
    \includegraphics[width=\linewidth]{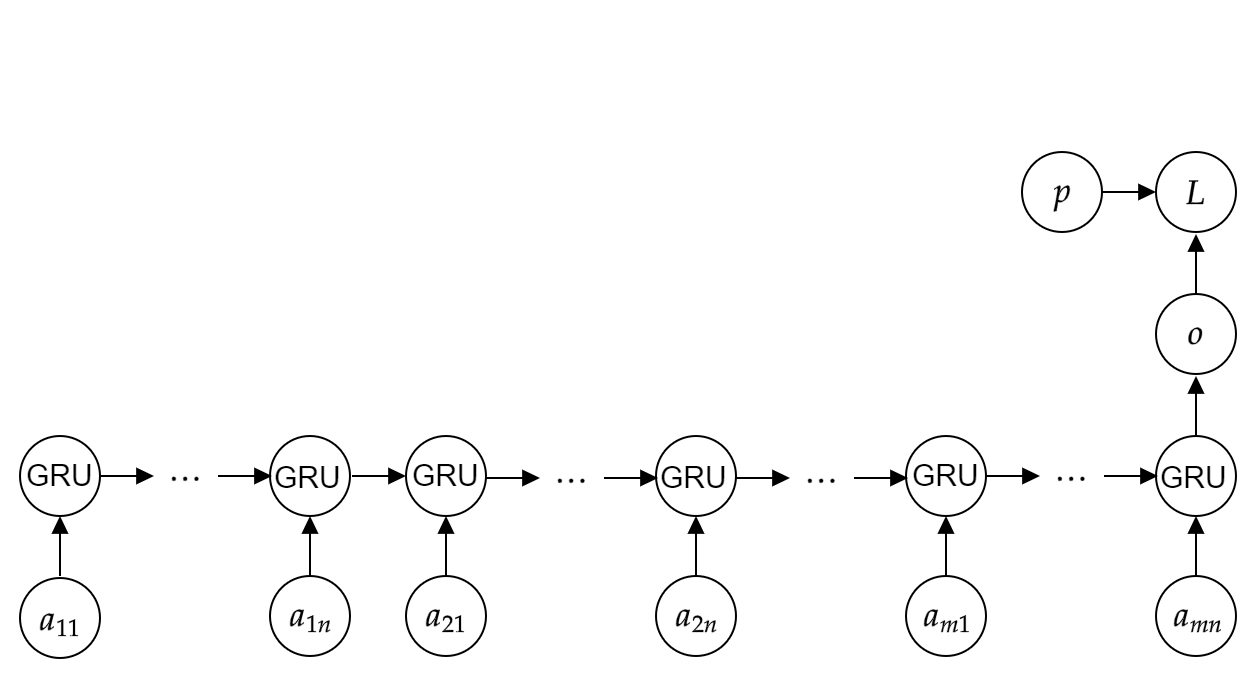}
    \caption{Architecture of cross-sessions concat}
    \label{fig:architecture_concat}
\end{minipage}%
\begin{minipage}[t]{.33\textwidth}
    \includegraphics[width=\linewidth]{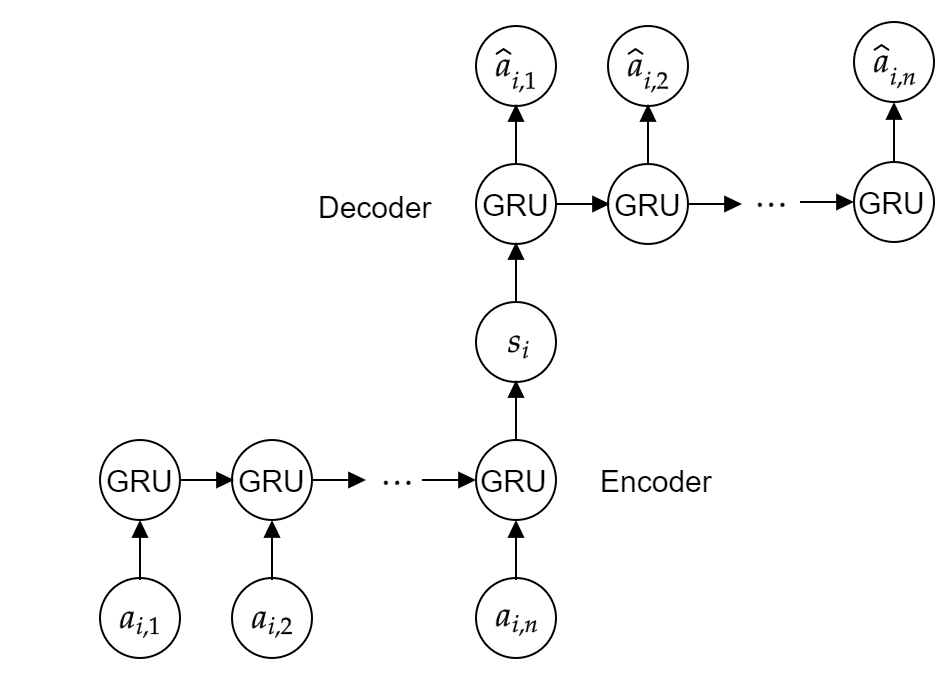}
    \caption{Architecture of cross-sessions auto}
    \label{fig:architecture_auto}
\end{minipage}
\end{figure*}

Our model is inspired by GRU4REC~\cite{HidasiEtAl2016}, an RNN with a single GRU~\cite{ChoEtAl2014} layer that models user interactions with items in a single session. The RNN has as input the ordered sequence of items interacted with in the session, and outputs for every time step the likelihood of each item being the item that the user interacts with next. Our cross-sessions RS extends GRU4REC by (1) taking multiple sessions of each user as input, as in Eq.~\eqref{eq:task}; (2) using various types of input actions that are not always associated with items; (3) predicting what items the user will buy after the last time step as opposed to predicting the next interaction for every time step in the sequence. 
Next we explain different ways of passing input sessions through the RNN. 

In the first way, which we call \emph{Cross-sessions Encode} (see Fig.~\ref{fig:architecture_encode}), we encode a session by aggregating over the actions in the session with a max pooling operation:
\begin{equation}
    s_i = \text{max}_{element}(a_{i1},a_{i2},a_{i3},...,a_{in}),
    \label{eqn:encoding}
\end{equation}
where $a_{ij}$ is the binarized vector marking the presence of an action section, action object and action type performed by a user at time step $j$ in session $i$, and $\max_{element}(\cdot)$ is a function that takes the element-wise maximum of vectors. Then, for every time step $i$ in the sequence of a user's sessions, an RNN with a single GRU layer computes the hidden state
\begin{align}
  \begin{aligned}
   h_i &= (1-z_i) \cdot h_{i-1} + z_i \cdot \hat{h}_i \\
   z_i &= \sigma(W_z s_i+U_z h_{i-1}), \\
   \hat{h}_i &= \tanh(W s_i+U(r_i \cdot h_{i-1})), \\
   r_i &= \sigma(W_r s_i + U_r h_{i-1}),
  \end{aligned}
  &&
  \begin{aligned}
   &\text{for}~~ i=1,..,m, \\ 
   &(update~~gate) \\
   &(candidate~~gate) \\
   &(reset~~gate)
  \end{aligned}
  \label{eqn:hidden state}
\end{align}
where $W_z, U_z, W, U, W_r$ and $U_r$ are weight matrices and $\sigma(\cdot)$ is the sigmoid function. The reset gate plays the role of forgetting information about the past that is not important given the current session. The update gate plays the role of judging whether the current session contains relevant information that should be stored. The hidden state, $h_i$, is a linear interpolation between the previous hidden state and the candidate gate.

Our second way of passing input sessions through the RNN is called \emph{Cross-sessions Concat} (see Fig.~\ref{fig:architecture_concat}). Here, we concatenate all sessions of a user into a single session $s=\lbrace a_{11},..,a_{1 n},a_{21},..,$ $a_{2 n},..,a_{m1},..,a_{m n}\rbrace$. Now the hidden state in Eq.~\eqref{eqn:hidden state} is computed for every time step $(ij)$ in $s$. Thereby this approach takes into account the overall order of actions across sessions, whereas the cross-sessions encode only accounts for the order of sessions.

In both cases, the RNN returns an output vector, $o$, of length $K$ after the last time step. Because a user can buy multiple items at the same time, we consider the learning task as multi-label classification and use the sigmoid function, $\sigma(\cdot)$, on each element of $o$ as output activation function to compute the likelihood of purchase
\begin{equation}
    \hat{p}_k = \sigma(o_k), ~~\text{for}~~ k=1,\ldots,K.
\end{equation}
During training the loss function is computed by comparing $\hat{p}$ with the binarized vector of the items purchased, $p$. Due to the learning task being multi-label classification, we define the loss function as the sum of the binary cross entropy loss over all items. The loss function is thereby different from the ranking loss used in GRU4REC and is given by
\begin{equation}
    L = - \sum_{k=1}^K \Big( p_k \cdot \log (\hat{p}_k) +(1-p_k) \cdot \log (1-\hat{p}_k) \Big).
    \label{eq:loss_function}
\end{equation}

We also use another variation of session encoding, \emph{Cross-sessions Auto} (see Fig.~\ref{fig:architecture_auto}), which automatically learns encodings of sessions with an autoencoder, instead of Eq. (\ref{eqn:encoding}). We train an RNN-based autoencoder with a single GRU layer that takes as input the ordered sequence of actions in a session and is evaluated on recreating the input using categorical cross-entropy loss on each of the $3$ features: action section, action object and action type. Once trained, the encoder is used to encode a session into a single vector, $s_i$, that can be used as input for Eq.~\eqref{eqn:hidden state}. The architecture of the autoencoder in Fig.~\ref{fig:architecture_auto} is then combined with the architecture for the recommendation part in Fig. \ref{fig:architecture_encode}.

\begin{figure}[tb]
    \centering
    \includegraphics[width=0.25\textwidth]{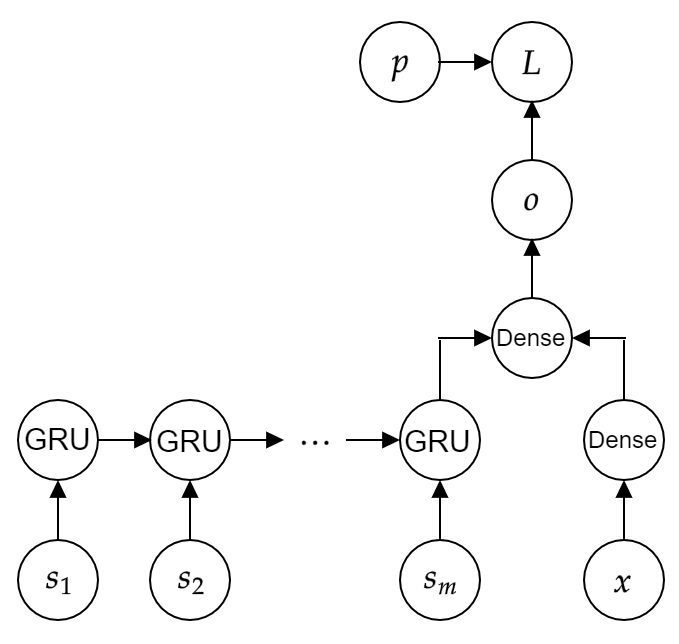}
    \caption{Architecture of a hybrid model between a cross-sessions and a demographic model. $x$ denotes an input vector representing demographic features of the user.}
    \label{fig:architecture - hybrid}
\end{figure}

Finally, we use a hybrid of a cross-sessions and a demographic model where the hidden state from a cross-sessions RNN is merged with the hidden state (a dense layer) from a feed forward neural network with demographic input features of the user. The concatenation of the two is passed through a dense layer. The architecture of Cross-sessions Encode combined with demographics is illustrated in Fig. \ref{fig:architecture - hybrid}.
The loss functions is cross entropy as in Eq.~\eqref{eq:loss_function}.

\section{Dataset}
\label{sec:dataset}

To the best of our knowledge, there exists no publicly available dataset that satisfies the criteria of our setup, namely: (1) item scarcity, (2) target action happening outside the session, and (3) actions of several types that are not always associated with items. We use a real-life dataset that we have obtained from a commercial insurance vendor and that we make freely available\footnote{\url{https://github.com/simonebbruun/cross-sessions_RS}} to the research community.
Next, we describe this dataset.

\subsection{Dataset Description}
\label{subsec:dataset}

\begin{table}[tb]
\parbox[t]{.45\textwidth}{
\centering
        \caption{Main properties of the dataset (*mean/std).}
        \begin{tabular}{@{}lr@{}}
            \toprule
            Users                        & 44,434     \\
            Items                        & 16         \\
            Purchase events              & 53,757     \\
            Sessions                     & 117,163    \\
            Actions                      & 1,256,156  \\ \midrule
            Purchase events per user*    & 1.21/0.51  \\
            Sessions before purchase event* & 2.18/1.68  \\
            Actions per session*         & 10.72/7.85 \\ \bottomrule
        \end{tabular}
        \label{tab:Dataset}
}
\hfill
\parbox[t]{.45\textwidth}{
     \centering
        \caption{Amount of different actions in the dataset.}
        \begin{tabular}{@{}llr@{}}
            \toprule
            \multirow{4}{*}{Action section} & E-commerce       & 256,319 (20.41\%) \\
                                            & Claims reporting & 11,188 (0.89\%)   \\
                                            & Information      & 198,147 (15.77\%) \\
                                            & Personal account & 790,502 (62.93\%) \\ \midrule
            \multirow{3}{*}{Action object}  & Items            & 249,378 (19.85\%) \\
                                            & Services         & 655,574 (52.19\%) \\
                                            & No object        & 351,204 (27.96\%) \\ \midrule
            \multirow{4}{*}{Action type}    & Click            & 811,747 (64.62\%) \\
                                            & Start            & 388,215 (30.90\%) \\
                                            & Act              & 47,713 (3.80\%)   \\
                                            & Complete         & 8,481 (0.68\%)    \\ \bottomrule
        \end{tabular}
        \label{tab:actions}
}
\end{table}

Tab.~\ref{tab:Dataset} shows general statistics of the dataset that was collected from the website of an insurance vendor between October 1, 2018 to September 30, 2020.
During this time, there was no RS running on the website that could influence the behaviour of users. We collected \emph{purchase events} of insurance products and additional coverages made by existing customers.
Customers were identified through log-in and cookies.
Both purchases made on the website and over the phone are included. $75\%$ of the users navigate through the company website, but still prefer to make the purchase over the phone.
This resulted in $53$K purchases corresponding to $44$K users. We observe no major seasonal effects in the purchase frequency.

For each user in the dataset, we collected all \emph{sessions} that occurred before the user's purchase event. 
A session consists of user actions on the website. 
The actions come with timestamps, action section (e-commerce, claims reporting, information, and personal account), action object (items and services), and action type (click, start, act, and complete).
Tab.~\ref{tab:actions} shows all the actions and their frequency in the dataset.
Most actions occur on the personal account page ($63\%$) and the e-commerce section ($20\%$) because the sessions are generated by existing customers in the period before they make a new purchase.
Users interact mainly with services ($52\%$).
Example of services are specification of ``employment'' (required if you have an accident insurance), specification of ``annual mileage'' (required if you have a car insurance), and information about ``the insurance process when moving home''.
Users can also click on the different sections without interacting with any objects, we denote this with ``no object''.
Not surprisingly, most of the actions are clicks ($65\%$) or start ($31\%$).
Examples of type ``act'' are ``change'', e.g., employment or ``fill out'' claims report. The action type ``complete'' can occur when a user completes a change, of e.g., employment, or completes a filled out claims report.

The dataset includes also \emph{demographic attributes} and \emph{portfolios} of each user. 
Demographic attributes include age, employment, income, residence, marital status, children, etc.
These features are aggregated at the address level, i.e., they represent the average across users living in the same area.
Portfolios represent the user history within the insurance company, e.g., items that the user has already bought.
Note that our cross-sessions models do not exploit demographic attributes or portfolios, but these are needed by state-of-the-art RSs we consider as baselines (see Section~\ref{subsec:exp_set_up}).

This dataset is different from publicly available datasets for session-based recommender tasks, e.g., Last.fm, RecSys Challenge (RSC) datasets, etc. in the following ways.
There is scarcity both in terms of items and user interactions.
The total number of items is $16$, much smaller than usual item sets, for example RSC15 contains $29$K items and Last.fm $91$K items.
In terms of user interactions, there are only $1.2$ purchase and $2.2$ sessions per users over a period of time spanning $2$ years.
Motivated by the scarcity of user interactions, we decided to log all types of user actions.
Therefore, we do not include only clicks and purchases/views, as in most of the available datasets, but we also record the whole user experience through the company website.

\subsection{Estimation of Session Threshold}
\label{subsec:threshold}

A common challenge when mining log data is how to determine the session length~\cite{JansenEtAl2006,JansenEtAl2007}: we can record the start timestamp of a user session but we cannot always record the end timestamp, for example when a user leaves the browser window open without logging out.
We face a similar problem but with respect to the length of the list of sessions instead of the length of a single session, i.e., we need to determine the amount of time such that two sessions can be grouped under the same task.
Specifically, we need to estimate the threshold $t$, such that given two subsequent sessions $s_1$ and $s_2$ from the same user, if $\mathtt{start\_time}(s_2) - \mathtt{start\_time}(s_1) \leq t$, then $s_1$ and $s_2$ belong to the same task. 
This threshold will determine the maximum number of sessions to be considered by our RSs.

A rule of thumb for Web sessions is to set the session end after $30$ minutes of inactivity, but a similar estimate does not exist with respect to inter sessions inactivity time.
\citet{HalfakerEtAl2015} estimate when a session ends directly from log data, i.e., as the amount of inactivity time that allows to deem a session as concluded.
We follow a similar approach applied on the whole list of sessions instead of actions within a single session to estimate when a task ends.

\begin{figure}[tb]
    \centering
    \includegraphics[width=0.4\textwidth]{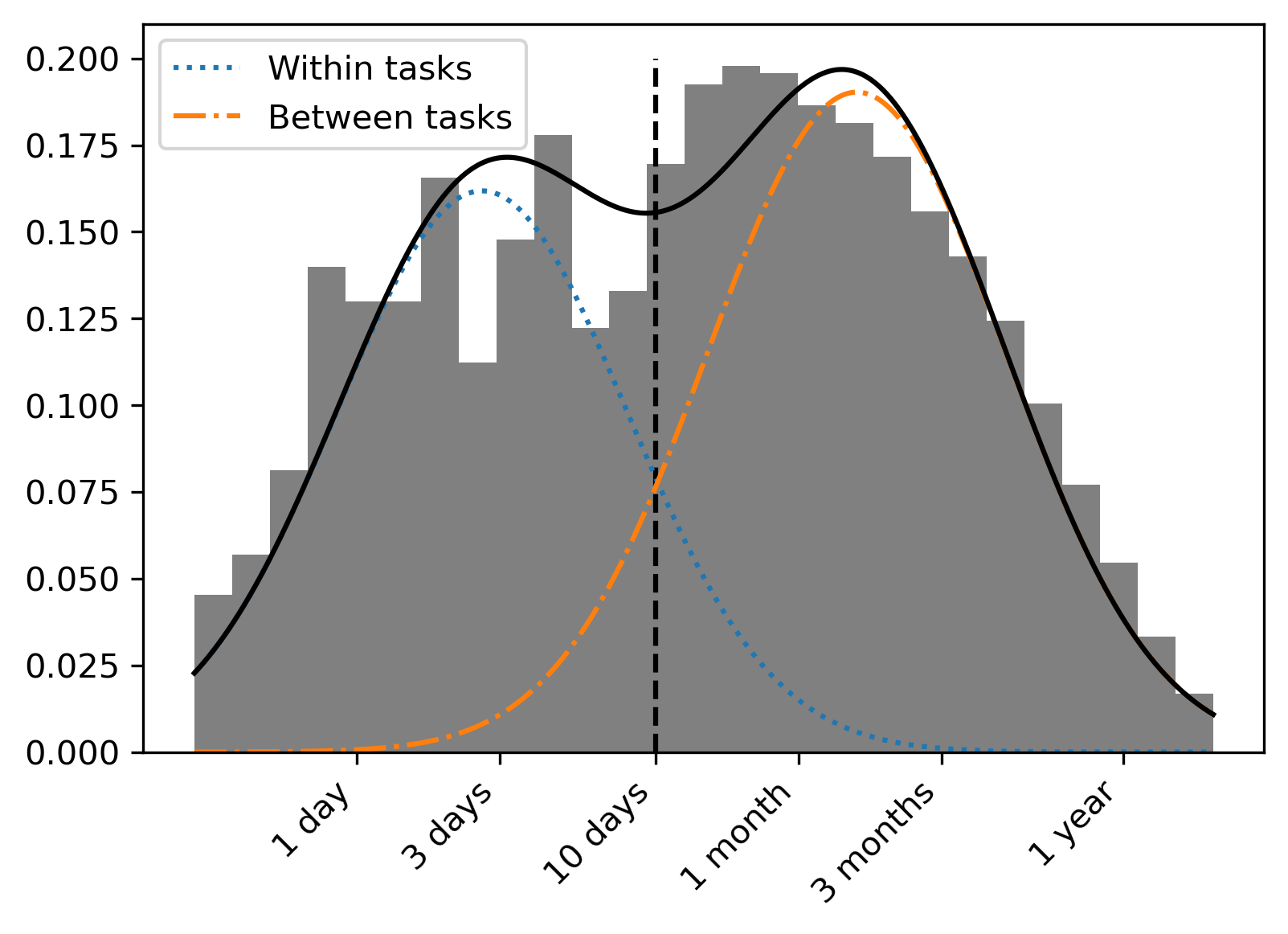}
    \caption{Histogram of logarithmically scaled inter-session times and fitted Gaussian mixture model.}
    \label{fig:inter-session times}
\end{figure}

First, we compute inter-session times, that is the times between consecutive sessions by the same user.
Then we plot the logarithmically scaled histogram of inter sessions times (see Fig.~\ref{fig:inter-session times}).
As reported in~\cite{HalfakerEtAl2015}, we observe a bimodal distribution with a valley.
Therefore we fit a two components Gaussian mixture model using Expectation Maximization and assume that the inter-session times are a mixture of times between: (1) sessions within the same task (blue line) and, (2) sessions belonging to different tasks (orange line). 
We set the inactivity threshold $t$ equal to the point where an inter-session time is equally likely to belong to one of the two distributions.
As illustrated in Fig.~\ref{fig:inter-session times}, the resulting threshold
is $10$ days. Thus two sessions belong to the same task if there is no longer than $10$ days between them.

\subsection{Dataset Pre-processing}
\label{subsec:preprocessing}

We pre-process the data as described next. All statistics in Tab.~\ref{tab:Dataset} and~\ref{tab:actions} refer to the pre-processed dataset.

From the purchase events we remove items that have frequency $<0.1\%$ and from the actions we remove sections, objects and types that have frequency $<0.1\%$, since low frequency items and actions are not optimal for modeling.
Consecutive repeated actions of the same kind are discarded, because they very likely represent noise, e.g., a user clicking twice due to latency from the website.
All sessions with $<3$ actions are removed, as they are poorly informative sessions.
All sessions are truncated in the end to have maximum $30$ actions (the $95$th percentile), to avoid very long training times. 
All users are kept, even those that have only a single recent session. Lastly, the $10$ days threshold estimated in Section~\ref{subsec:threshold} motivates two additional pre-processing steps.
First, sessions that exceed the $10$ days threshold are discarded.
Second, within the $10$ days rule, the list of recent sessions for each user is truncated to maximum $7$ sessions (the $95$th percentile).

\section{Experiments}
\label{sec:experiments}
We present the experimental set-up (Section~\ref{subsec:exp_set_up}) and the results (Section~\ref{subsec:results}).
Our source code is publicly available\footnote{\url{https://github.com/simonebbruun/cross-sessions_RS}}.

\subsection{Experimental Set-up}
\label{subsec:exp_set_up}
First we describe the evaluation procedure, then the baselines, implementation details and hyperparameter tuning.

\paragraph{Evaluation Procedure}
As test set, we use the latest $10\%$ of purchase events with associated past sessions. The remaining $90\%$ is used for training. 
Since some users have had multiple purchase events, we remove purchase events from the training data, if their associated past sessions also appear in the test set. 

The RS generates a prediction for what items the user will buy as a ranked list, i.e., items are ranked by their predicted probability: the closer to the top of the ranking, the higher the estimated probability of the item. There are two types of items: new insurance products and additional coverage.
Since it is only possible for a user to buy an additional coverage if the user has the corresponding base insurance product, we use a post filter to set the probability to $0$ if that is not the case, as per~\cite{Aggarwal2016}.
The resulting list of ranked items is evaluated with Hit Rate (HR), Mean Reciprocal Rank (MRR), Precision, Recall, and Mean Average Precision (MAP).
We use a cutoff threshold of $3$ because: (1) the total number of items is $16$, therefore high cut-offs, e.g., $\geq10$, will increase recall and all measures will reach high values, which will not inform on the actual quality of the RSs; (2) on the user interface the user will be recommended up to $3$ items.
Additionally, we report HR and MRR scores for all cut-offs value from $1$ to $5$.

Experimental results are supported by statistical testing. For HR we use McNemar's test \cite{Dietterich1998} and for all other measure we use one-way ANOVA \cite{kutner2005}, both with a confidence level of $0.05$, and post hoc tests to control the family-wise error rate due to multiple comparisons.

\paragraph{Baselines} We compare our models against the following state-of-the-art baselines:
\begin{itemize}
    \item \textbf{Random} recommends random items to the user.
    \item \textbf{Popular} recommends the items with the largest number of purchases across users.
    \item \textbf{SVD} is a method that factorizes the user-item matrix by singular value decomposition~\cite{CremonesiEtAl2010}. The portfolio data forms the user-item matrix, where a user-item entry is $1$ if the user has bought the item and $0$ otherwise. Since in the insurance domain a user can buy the same item again (e.g., a second car insurance) and matrix factorization cannot be used for repeated recommendations, we add repeated items as new items (columns) to the matrix.
    \item \textbf{Demographic} is a classification model, as per~\cite{Qazi2017, Qazi2020}, that uses user demographics and their portfolios as input features. The portfolios are represented with a feature for each item counting how many of the items the user has already bought. Demographic features and portfolio features are concatenated. We use a feed forward neural network to make a fair comparison with the neural session-based approaches.
    \item \textbf{GRU4REC} is a neural session-based model~\cite{HidasiEtAl2016,HidasiAndKaratzoglou2018}. Its input is the last session of a user, consisting of the 3-tuple user actions described in Section \ref{subsec:approach}. For every time step, the model outputs the likelihood for each action to be the one the user interacts with next.  Recommendations are based on the output after the final time step.
    \item \textbf{GRU4REC Concat} is the same as GRU4REC, but all recent user sessions are concatenated into a single session.
    \item \textbf{SKNN\_E} is a session-based nearest neighbour model as the one presented in~\cite{JannachAndLudewig2017} with the extension suggested in~\cite{Latifi2021}. The nearest neighbours are determined based on the set of actions in all recent sessions of each user. A user's set of actions is a vector computed with a max pooling operation over all actions generated by the user (in recent sessions). We then adapt this baseline to our task, so the recommendations are based on the items purchased by the neighbours of the target user rather than the items interacted with in the ongoing session.
    \item \textbf{SKNN\_EB} is the same as SKNN\_E, but with a further extension suggested in~\cite{Latifi2021}: scores of items previously interacted with are boosted with a factor. The factor is tuned as a hyperparameter.
\end{itemize}
Note that SVD and the demographic model make use of user portfolio, i.e., users past purchases, while this is not the case for all cross-sessions models, GRU4REC and SKNN. GRU4REC is included as it has shown best performance under identical conditions on various datasets among all the neural models compared in~\cite{LudewigEtAl2021} and in~\cite{Latifi2021}.

We tried the sequential extension to SKNN, Vector Multiplication SKNN, which is presented in~\cite{LudewigAndJannach2018}, but did not obtain better performance than the original one. Sequence and Time Aware Neighborhood~\cite{GargEtAL2019} is not included as baseline since it was not possible to adapt it to the task under consideration (for the reasons discussed in Section~\ref{subsec:related_session}).

\paragraph{Implementation \& Hyperparameters}

All implementation is in \texttt{Python} \texttt{3.7.4} and \texttt{Ten\-sor\-Flow} \texttt{2.6.0}\footnote{We used Tensorflow's implementation of padding and masking to deal with variable length input in the RNNs.}. We used \texttt{Adam} as the optimizer with TensorFlow's default settings for the learning rate, exponential decay rates and the epsilon parameter. Early stopping was used to choose the number of epochs based on the minimum loss on the validation dataset. We used two-layer networks\footnote{In all models the second layer is a dense layer with ReLU activation function.} with dropout regularization on the first hidden layer.

We partitioned the training set in the same way as the whole dataset, so the validation set includes the latest $10\%$ of purchases with associated sessions and the remaining is used for training. 
We tuned the hyperparameters of each neural model (batch size, number of units, and dropout rate) on the validation set using grid search. We test powers of $2$ for the batch size and number of units ranging from $16$ to $512$. For the dropout rate, we test values in $[0.1,0.5]$ with step size $0.1$.
The final hyperparameters used are reported in Tab.~\ref{tab:hyperparameters}.

For GRU4REC and GRU4REC Concat we tried the 3 different loss functions: cross-entropy, BPR \cite{Rendle2009} and TOP1 \cite{HidasiEtAl2016}. Cross-entropy was finally chosen for both models, as it performed best on the validation set. For the non-neural models, the optimal number of latent factors for the SVD model was $1$, the optimal number of neighbours for both SKNN\_E and SKNN\_EB was $30$, and the optimal boost factor for SKNN\_EB was $0.5$. Neural models were trained on Nvidia GeForce MX250 equipped with 2GB of GPU memory. The maximum training time was ~$6$ hours.

\begin{table}[tb]
\caption{Hyperparameters (*autoencoder/RNN)}
\resizebox{0.4\textwidth}{!}{%
\begin{tabular}{@{}lrrr@{}}
\toprule
\multicolumn{1}{c}{Model} & \multicolumn{1}{c}{Batch size} & \multicolumn{1}{c}{Units} & \multicolumn{1}{c}{Dropout} \\ \midrule
Demographic               & 32                             & 32                        & 0.3                         \\
GRU4REC                   & 32                             & 256                       & 0.2                         \\
GRU4REC Concat            & 32                             & 256                       & 0.2                         \\
Cross-sessions Concat     & 128                            & 64                        & 0.3                         \\
Cross-sessions Encode     & 32                             & 64                        & 0.3                         \\
Cross-sessions Auto*      & 128/32                         & 512/64                    & -/0.4                       \\ \bottomrule
\end{tabular}
}%
\label{tab:hyperparameters}
\end{table}

\subsection{Experimental Results}
\label{subsec:results}

Next we compare our cross-sessions against state-of-the-art baselines.
Furthermore, we combine cross-sessions models with the demographic model.
We further breakdown the analysis of our model to understand the impact of exploiting all past sessions and actions instead of only the most recent ones and their order. Lastly, we conduct an ablation study to show how different actions (sections, objects, and type) affect the performance of our models.

\begin{table*}[tb]
\caption{Performance results. All results marked with * are significantly different from cross-sessions encode.
The best score for each measure is in bold. Percentages in brackets denote the difference of our models from the strongest baseline (SKNN\_EB).}
\begin{tabular}{@{}lrrrrr@{}}
\toprule
\multicolumn{1}{c}{Model} & \multicolumn{1}{c}{HR@3} & \multicolumn{1}{c}{Precision@3} & \multicolumn{1}{c}{Recall@3} & \multicolumn{1}{c}{MRR@3} & \multicolumn{1}{c}{MAP@3} \\ \midrule
Random                    & 0.3235*                  & 0.1114*                         & 0.2940*                      & 0.1910*                   & 0.1839*                   \\
Popular                   & 0.6217*                  & 0.2145*                         & 0.5855*                      & 0.4764*                   & 0.4540*                   \\
SVD                       & 0.6646*                  & 0.2372*                         & 0.6327*                      & 0.4997*                   & 0.4829*                   \\
Demographic               & 0.7392*                  & 0.2649*                         & 0.7095*                      & 0.5620*                   & 0.5446*                   \\
GRU4REC                   & 0.6479*                  & 0.2313*                         & 0.6208*                      & 0.5443*                   & 0.5264*                   \\
GRU4REC Concat            & 0.6616*                  & 0.2365*                         & 0.6362*                      & 0.5620*                   & 0.5453*                   \\
SKNN\_E                   & 0.8106*                  & 0.2914*                         & 0.7848*                      & 0.6740*                   & 0.6567*                   \\
SKNN\_EB                  & 0.8132*                  & 0.2922*                         & 0.7872*                      & 0.6785*                   & 0.6610*                   \\ \midrule
Cross-sessions Encode     & \textbf{0.8380 (3.04\%)} & \textbf{0.3030 (3.67\%)}        & \textbf{0.8145 (3.46\%)}     & \textbf{0.7093 (4.53\%)}  & \textbf{0.6923 (4.73\%)}  \\
Cross-sessions Concat     & 0.8265 (1.62\%)          & 0.2984 (2.12\%)                 & 0.8019 (1.87\%)              & 0.7051 (3.92\%)           & 0.6876 (4.02\%)           \\
Cross-sessions Auto       & 0.8356 (2.74\%)          & 0.3024 (3.48\%)                 & 0.8128 (3.24\%)              & 0.7085 (4.41\%)           & 0.692 (4.69\%)            \\ \bottomrule
\end{tabular}
\label{tab:results}
\end{table*}

\begin{table*}[tb]
\caption{The versions of our models enhanced with demographic data. The rest of notation is as in Tab.~\ref{tab:results}.}
\resizebox{\textwidth}{!}{%
\begin{tabular}{@{}lrrrrr@{}}
\toprule
\multicolumn{1}{c}{Model}              & \multicolumn{1}{c}{HR@3}  & \multicolumn{1}{c}{Precision@3} & \multicolumn{1}{c}{Recall@3} & \multicolumn{1}{c}{MRR@3} & \multicolumn{1}{c}{MAP@3} \\ \midrule
Cross-sessions Encode with Demographic & \textbf{0.8542* (5.03\%)} & \textbf{0.3103* (6.17\%)}       & \textbf{0.8313* (5.6\%)}     & 0.7268* (7.11\%)          & 0.7099* (7.41\%)          \\
Cross-sessions Concat with Demographic & 0.8497* (4.48\%)          & 0.3087* (5.64\%)                & 0.8269* (5.04\%)             & \textbf{0.7298* (7.55\%)} & \textbf{0.7131* (7.88\%)} \\
Cross-sessions Auto with Demographic   & 0.8460 (4.03\%)           & 0.3072 (5.13\%)                 & 0.8228 (4.52\%)              & 0.7223 (6.45\%)           & 0.7050 (6.66\%)           \\ \bottomrule
\end{tabular}
}%
\label{tab:results2}
\end{table*}

\begin{figure}[tb]
    \centering
    \begin{subfigure}{0.4\textwidth}
        \centering
        \includegraphics[width=\columnwidth]{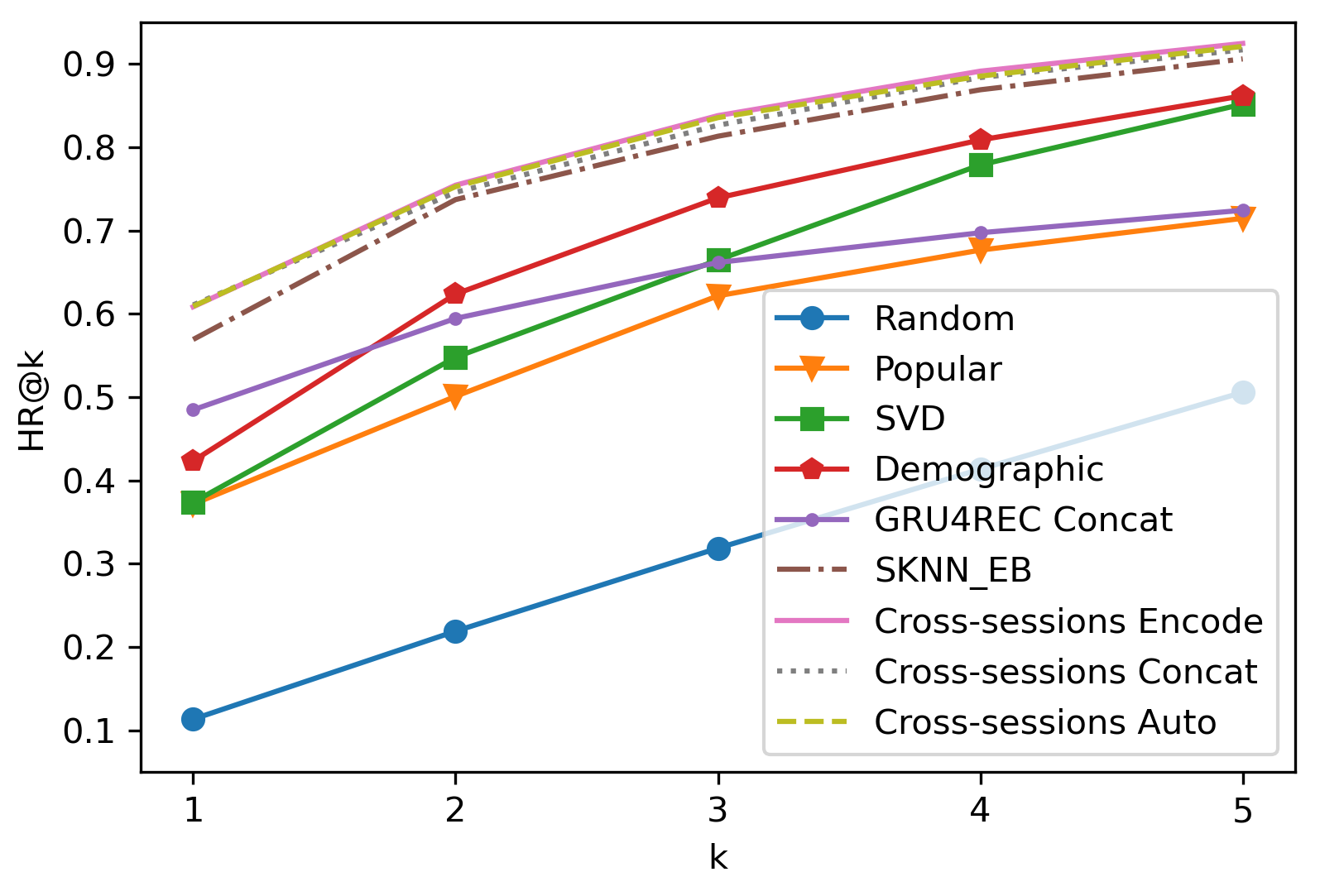}
    \end{subfigure}
    \begin{subfigure}{0.4\textwidth}
        \centering
        \includegraphics[width=\columnwidth]{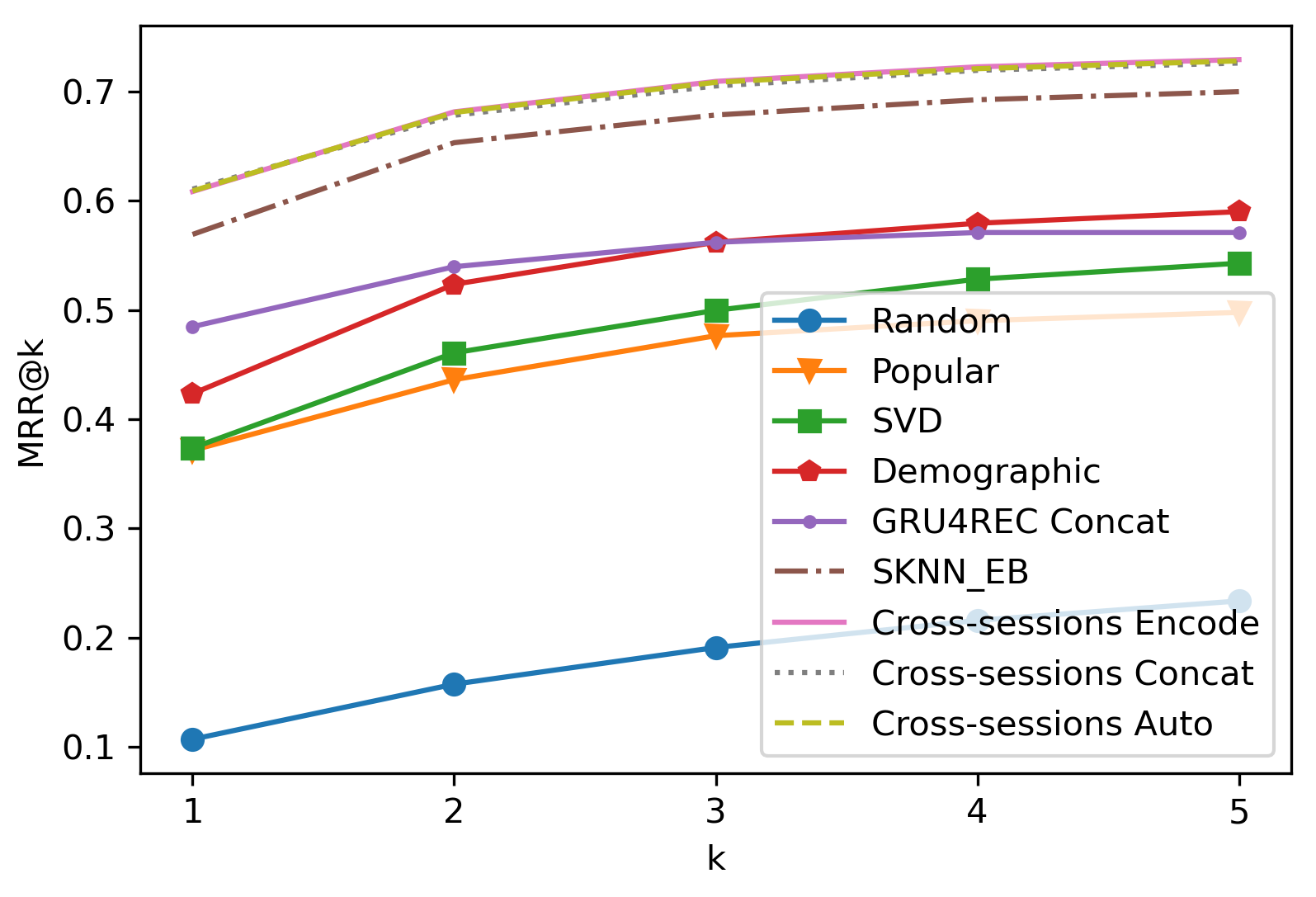}
    \end{subfigure}
    \caption{HR@k and MRR@k for varying choices of the cutoff threshold $k$.}
    \label{fig:varying cutoffs}
\end{figure}

\begin{figure}[tb]
    \centering
    \begin{subfigure}{0.4\textwidth}
        \centering
        \includegraphics[width=\columnwidth]{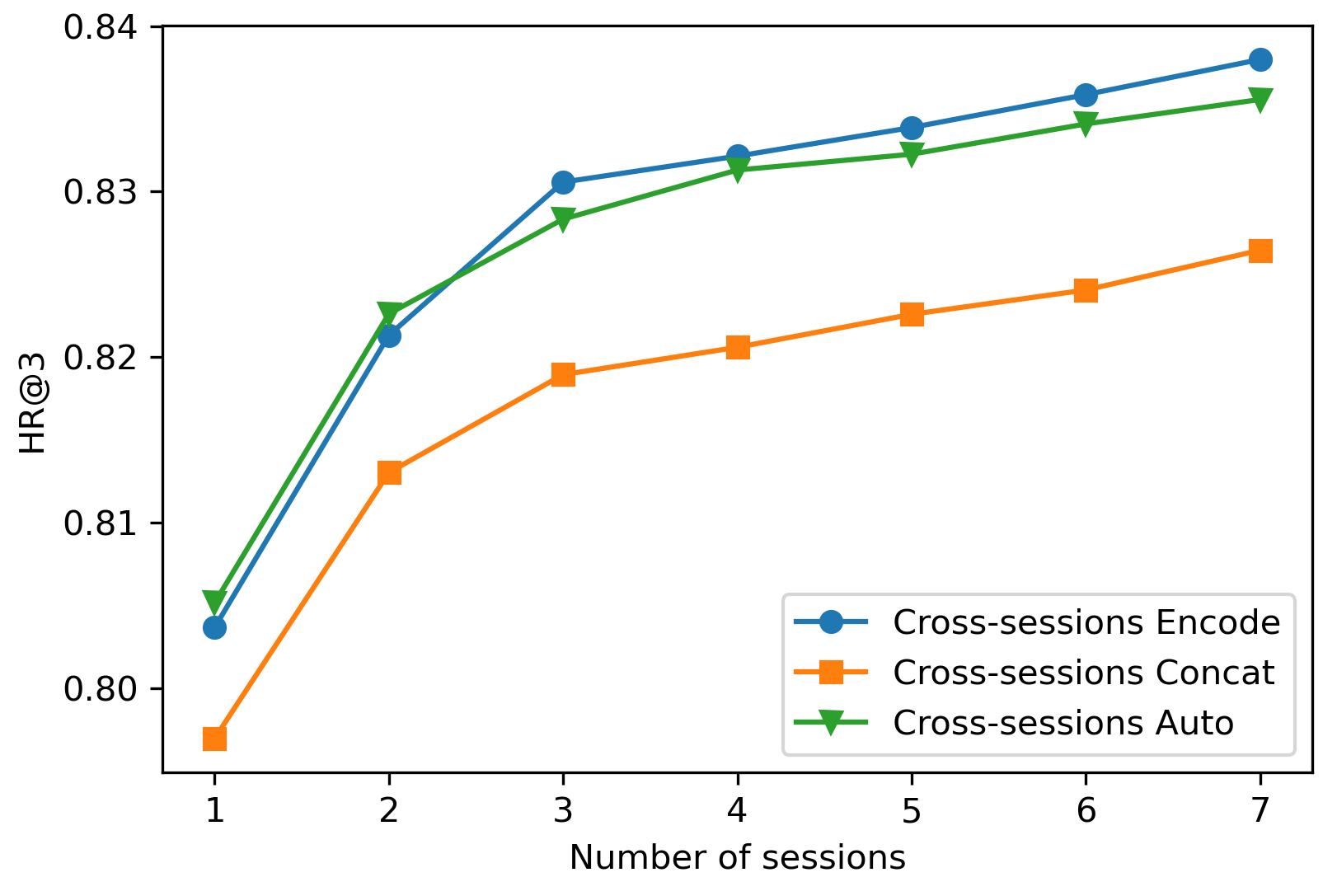}
    \end{subfigure}
    \begin{subfigure}{0.4\textwidth}
        \centering
        \includegraphics[width=\columnwidth]{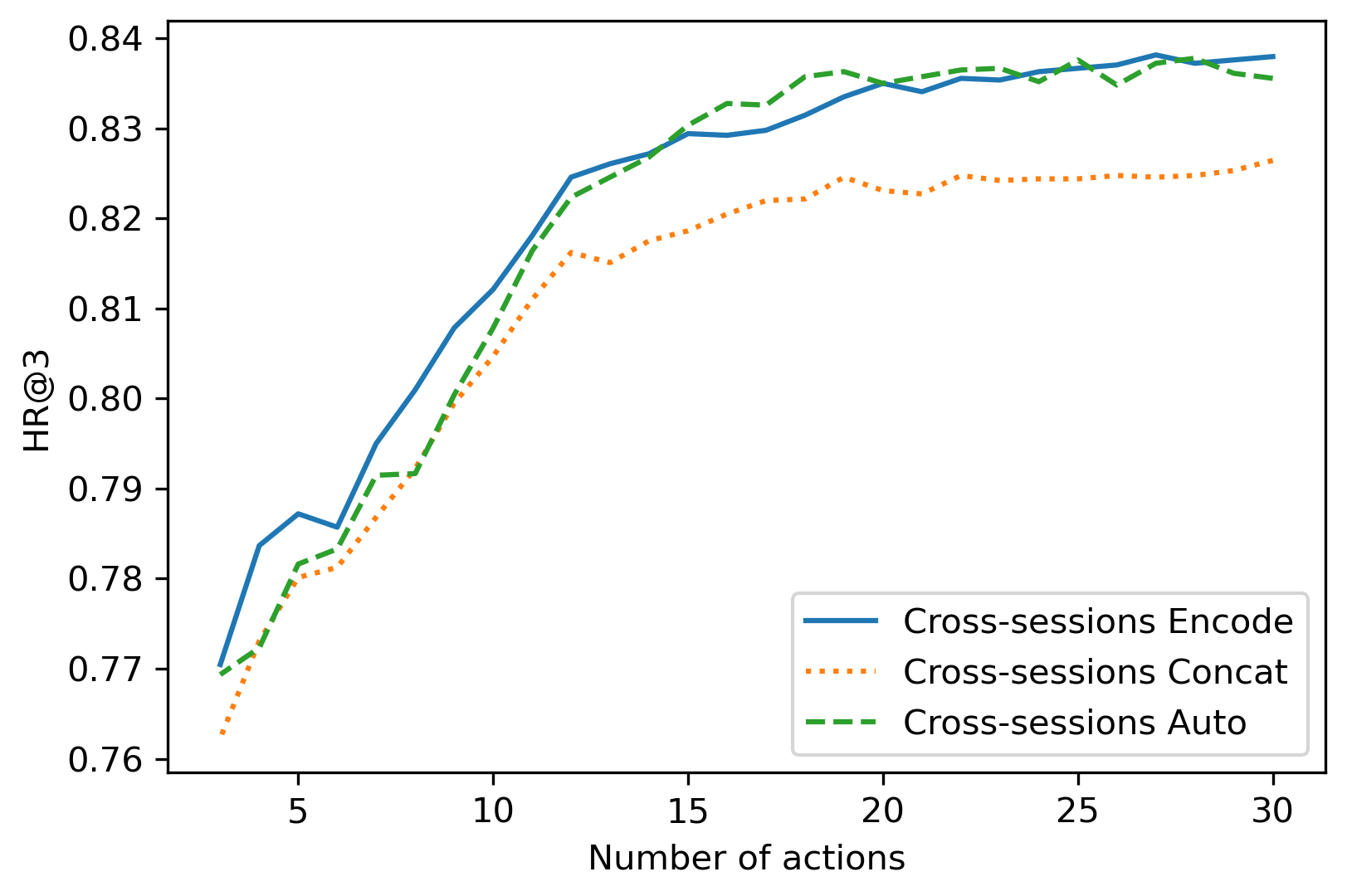}
    \end{subfigure}
    \caption{HR@3 for different number of recent sessions and  actions.}
    \label{fig:number of sessions}
\end{figure}

\paragraph{Performance Analysis}
Tab.~\ref{tab:results} compares our cross-sessions models against the baselines. 
The simple Popular model is a quite strong baseline, unlike in domains like retail and video services~\cite{HidasiEtAl2016,DacremaEtAl2021}, because of the few different items and the role of the post filter to make sure not to recommend items that the user cannot buy.

As in prior work on insurance RSs~\cite{Qazi2017}, we also see a significant improvement in using a demographic RS compared to the traditional matrix factorisation method, SVD, due to the sparse feedback on items in the insurance domain (see Tab.~\ref{tab:Dataset}) and users' demographic characteristics being good signals for learning insurance recommendations.

The session-based methods, SKNN\_E, SKNN\_EB and cross-sessions, significantly outperform the non-session-based methods, while this is not the case for GRU4REC and GRU4REC Concat. This shows that users' recent sessions are stronger signals for learning insurance recommendations than long-term preferences and demographic characteristics, but recommending the item that the user is most likely to interact with next on the website is not appropriate for insurance recommendations.
All cross-sessions methods significantly outperform SKNN\_E and SKNN\_EB suggesting that an RNN is better at modeling relationships between the user actions that lead to the purchase of specific items.

The results suggest that encoding of sessions is better than the trivial concatenation of sessions indicating that dependencies across sessions are important.
The results further suggest that the encoding of sessions with a max-pooling operation is better than the automatically learned encodings of sessions (using an autoencoder). This is most likely because the order of actions (which the autoencoder takes into account) adds more noise to the model than signal, or the autoencoder needs a larger amount of training data in order to effectively learn to encode sessions.

Fig.~\ref{fig:varying cutoffs} shows HR and MRR at varying cutoffs $k$ from $1$ to $5$.
We have similar results for recall, precision, and MAP, which are omitted due to space limitations.
The results are consistent over varying choices of the cutoff threshold, with the exception of GRU4REC Concat which is better than SVD and Demographic for smaller cutoff thresholds (1-2), but not for larger. Across all choices of the cutoff threshold there is a clear gap between cross-sessions models and the others.
The general trend for both measures is that they tend to increase as the cut-off $k$ increases. As expected, this happens because, by increasing the cutoff threshold, it is more likely to include the purchased items.

Finally, we combine our cross-sessions models with the demographics.
The results are shown in Tab.~\ref{tab:results2}.
The hybrid approach between a cross-sessions and a demographic model yields better performance than the individual models for all models and evaluation measures.
This indicates that the two types of information, sessions and demographic, capture different aspects of the problem.
The best results are obtained with \emph{Cross-sessions Encode} and \emph{Cross-sessions Concat} and both models are statistically significantly different from the \emph{Cross-sessions Encode} model without demographic data.

\paragraph{Analysis of Number of Sessions and Actions}

Next, we analyse how the number of sessions affects our cross-sessions models.
We break down the performance scores of our models based on the number of sessions, starting with only the most recent session, up to including all the available sessions (the maximum number of sessions per user is $7$, see Sections~\ref{subsec:threshold} and~\ref{subsec:preprocessing}).
Fig.~\ref{fig:number of sessions} (left) shows HR@$3$ computed for our cross-sessions models with respect to varying numbers of user sessions.
In general, there is an increasing trend in performance with the number of recent sessions, emphasising the additional contribution brought by all recent sessions of each user rather than just the last one.
We can observe that \emph{Cross-sessions Encode} and \emph{Cross-sessions Auto} consistently outperform \emph{Cross-sessions Concat} for each number of sessions.
We observe similar results for MRR, recall, precision, and MAP, which are not included here due to space limitations.

We do the same analysis with number of actions per session. Fig.~\ref{fig:number of sessions} (right) shows that HR@$3$ generally increases with the number of actions.
The growth is particularly steep up to about $10$ actions per session after which it flattens out.
We observe similar results for MRR, recall, precision, and MAP, which are not included here due to space limitations. 

\paragraph{Analysis of Session Order}
We analyse the importance of session order by randomly shuffling the order of sessions and retraining the models. We shuffle the order in both training, validation and test data, and perform the experiment $5$ times to account for randomness. The mean performance is presented in Tab.~\ref{tab:shuffled_session_order}. Across all our models and evaluation measures, performance drops when shuffling the session order, but the decrease is limited to less than $1\%$. The results indicate that the superiority of the cross-sessions models is not due to sequential dependencies, rather they are simply better at capturing the relationships between user actions and the purchase of specific items.

\begin{table*}[tb]
\caption{Study of session order. Relative change in parentheses.}
\resizebox{\textwidth}{!}{%
\begin{tabular}{@{}llrrrrr@{}}
\toprule
\multicolumn{2}{c}{Model}                      & \multicolumn{1}{c}{HR@3} & \multicolumn{1}{c}{Precision@3} & \multicolumn{1}{c}{Recall@3} & \multicolumn{1}{c}{MRR@3} & \multicolumn{1}{c}{MAP@3} \\ \midrule
Cross-sessions Encode & original session order & 0.8380                   & 0.3030                          & 0.8145                       & 0.7093                    & 0.6923                    \\
                      & shuffled session order & 0.8345 (-0.41\%)         & 0.3008 (-0.7\%)                 & 0.8096 (-0.60\%)             & 0.7058 (-0.49\%)          & 0.688 (-0.61\%)           \\ \midrule
Cross-sessions Concat & original session order & 0.8265                   & 0.2984                          & 0.8019                       & 0.7051                    & 0.6876                    \\
                      & shuffled session order & 0.8209 (-0.20\%)         & 0.2935 (-0.52\%)                & 0.7925 (-0.37\%)             & 0.6978 (-0.28\%)          & 0.6759 (-0.49\%)          \\ \midrule
Cross-sessions Auto   & original session order & 0.8356                   & 0.3024                          & 0.8128                       & 0.7085                    & 0.6920                    \\
                      & shuffled session order & 0.8305 (-0.61\%)         & 0.3003 (-0.70\%)                & 0.8069 (-0.72\%)             & 0.704 (-0.64\%)           & 0.6867 (-0.76\%)          \\ \bottomrule
\end{tabular}
}%
\label{tab:shuffled_session_order}
\end{table*}

\paragraph{Analysis of Actions}

\begin{table*}[tb]
\caption{Ablation study of actions. Relative change in parentheses.}
\resizebox{\textwidth}{!}{%
\begin{tabular}{@{}llrrrrr@{}}
\toprule
\multicolumn{2}{c}{Model}                                          & \multicolumn{1}{c}{HR@3} & \multicolumn{1}{c}{Precision@3} & \multicolumn{1}{c}{Recall@3} & \multicolumn{1}{c}{MRR@3} & \multicolumn{1}{c}{MAP@3} \\ \midrule
\multirow{10}{*}{Cross-sessions Encode} & all actions              & 0.8380                   & 0.3030                          & 0.8145                       & 0.7093                    & 0.6923                    \\
                                        & without E-commerce       & 0.7526 (-10.19\%)        & 0.2698 (-10.95\%)               & 0.7249 (-11.00\%)            & 0.5951 (-16.09\%)         & 0.5764   (-16.74\%)       \\
                                        & without Claims reporting & 0.8250 (-1.55\%)         & 0.2979 (-1.68\%)                & 0.8012 (-1.64\%)             & 0.7006 (-1.22\%)          & 0.6829 (-1.35\%)          \\
                                        & without Information      & 0.8317 (-0.75\%)         & 0.3000 (-0.98\%)                & 0.8072 (-0.89\%)             & 0.7045 (-0.68\%)          & 0.6863 (-0.87\%)          \\
                                        & without Personal account & 0.8067 (-3.73\%)         & 0.2894 (-4.48\%)                & 0.7803 (-4.19\%)             & 0.6604 (-6.89\%)          & 0.6438 (-7.00\%)          \\
                                        & without Items            & 0.7379 (-11.94\%)        & 0.2652 (-12.46\%)               & 0.7116 (-12.63\%)            & 0.5720 (-19.36\%)         & 0.5548 (-19.86\%)         \\
                                        & without Services         & 0.8032 (-4.15\%)         & 0.2880 (-4.93\%)                & 0.7765 (-4.66\%)             & 0.6639 (-6.40\%)          & 0.6465 (-6.61\%)          \\
                                        & without Start            & 0.8162 (-2.60\%)         & 0.2935 (-3.11\%)                & 0.7906 (-2.94\%)             & 0.6771 (-4.55\%)          & 0.6592 (-4.77\%)          \\
                                        & without Act              & 0.8318 (-0.73\%)         & 0.3005 (-0.80\%)                & 0.8082 (-0.77\%)             & 0.7035 (-0.82\%)          & 0.6864 (-0.85\%)          \\
                                        & without Complete         & 0.8317 (-0.75\%)         & 0.3005 (-0.82\%)                & 0.8078 (-0.82\%)             & 0.7036 (-0.80\%)          & 0.6861 (-0.89\%)          \\ \midrule
\multirow{10}{*}{Cross-sessions Concat} & all actions              & 0.8265                   & 0.2984                          & 0.8019                       & 0.7051                    & 0.6876                    \\
                                        & without E-commerce       & 0.7456   (-9.79\%)       & 0.2677 (-10.30\%)               & 0.7188 (-10.37\%)            & 0.5918 (-16.08\%)         & 0.5723   (-16.77\%)       \\
                                        & without Claims reporting & 0.8287 (+0.27\%)         & 0.2987 (+0.08\%)                & 0.8037 (+0.22\%)             & 0.7067 (+0.22\%)          & 0.6879 (0.04\%)           \\
                                        & without Information      & 0.8308 (+0.53\%)         & 0.3009 (+0.81\%)                & 0.8076 (+0.71\%)             & 0.7133 (+1.15\%)          & 0.6958 (1.19\%)           \\
                                        & without Personal account & 0.8130 (-1.63\%)         & 0.2914 (-2.35\%)                & 0.7872 (-1.84\%)             & 0.6795 (-3.64\%)          & 0.6633 (-3.52\%)          \\
                                        & without Items            & 0.7348 (-11.10\%)        & 0.2630 (-11.87\%)               & 0.7077 (-11.76\%)            & 0.5783 (-17.99\%)         & 0.5555 (-19.21\%)         \\
                                        & without Services         & 0.8119 (-1.76\%)         & 0.2914 (-2.37\%)                & 0.7856 (-2.04\%)             & 0.6798 (-3.59\%)          & 0.6635 (-3.50\%)          \\
                                        & without Start            & 0.8162 (-1.24\%)         & 0.2943 (-1.39\%)                & 0.7913 (-1.32\%)             & 0.6851 (-2.84\%)          & 0.6676 (-2.90\%)          \\
                                        & without Act              & 0.8304 (+0.47\%)         & 0.3001 (+0.56\%)                & 0.8063 (+0.55\%)             & 0.7083 (+0.44\%)          & 0.6908 (0.47\%)           \\
                                        & without Complete         & 0.8276 (+0.14\%)         & 0.2995 (+0.35\%)                & 0.8038 (+0.23\%)             & 0.7065 (+0.19\%)          & 0.6912 (0.91\%)           \\ \midrule
\multirow{10}{*}{Cross-sessions Auto}   & all actions              & 0.8356                   & 0.3024                          & 0.8128                       & 0.7085                    & 0.6920                    \\
                                        & without E-commerce       & 0.7473 (-10.57\%)        & 0.2691 (-11.02\%)               & 0.7206 (-11.34\%)            & 0.5953 (-15.97\%)         & 0.5776   (-16.53\%)       \\
                                        & without Claims reporting & 0.8227 (-1.54\%)         & 0.2966 (-1.93\%)                & 0.7979 (-1.83\%)             & 0.7001 (-1.19\%)          & 0.6819 (-1.45\%)          \\
                                        & without Information      & 0.8287 (-0.82\%)         & 0.2999 (-0.82\%)                & 0.8048 (-0.99\%)             & 0.7055 (-0.42\%)          & 0.6881 (-0.56\%)          \\
                                        & without Personal account & 0.8175 (-2.17\%)         & 0.2925 (-3.27\%)                & 0.7916 (-2.60\%)             & 0.6809 (-3.89\%)          & 0.6639 (-4.07\%)          \\
                                        & without Items            & 0.7398 (-11.46\%)        & 0.2649 (-12.41\%)               & 0.7110 (-12.52\%)            & 0.5771 (-18.55\%)         & 0.5585 (-19.29\%)         \\
                                        & without Services         & 0.8051 (-3.64\%)         & 0.2892 (-4.37\%)                & 0.7792 (-4.13\%)             & 0.6741 (-4.86\%)          & 0.6574 (-5.00\%)          \\
                                        & without Start            & 0.8183 (-2.07\%)         & 0.2942 (-2.71\%)                & 0.7924 (-2.50\%)             & 0.6842 (-3.43\%)          & 0.6661 (-3.74\%)          \\
                                        & without Act              & 0.8287 (-0.82\%)         & 0.2993 (-1.03\%)                & 0.8047 (-1.00\%)             & 0.6994 (-1.28\%)          & 0.6817 (-1.49\%)          \\
                                        & without Complete         & 0.8257 (-1.18\%)         & 0.2987 (-1.23\%)                & 0.8025 (-1.27\%)             & 0.7024 (-0.85\%)          & 0.6852 (-0.98\%)          \\ \bottomrule
\end{tabular}
}%
\label{tab:action_influence}
\end{table*}

We use ablation to analyse the influence of different actions, i.e., sections, objects, and types.
Each time we remove all actions of a given type and evaluate our cross-sessions models after re-training without the action type under analysis.
The results are presented in Tab.~\ref{tab:action_influence} for all our models and evaluation measures.
We did not consider the action type ``click'' in the ablation study because removing clicks results in removing most of the actions ($65\%$), but also most of the objects and sections, since users interact with objects and sections mainly through clicks.

For the \emph{Cross-sessions Encode} and \emph{Cross-sessions Auto} models, all actions contribute positively to the model performance, i.e., their removal causes a drop in the measure score.
We can have a similar conclusion for the \emph{Cross-sessions Concat} model with the exception that
the removal of actions in the section ``information'' and ``claims reporting'', or of type ``act'' and ``complete'' increases performance instead of decreasing it, but the actual difference is negligible (less than $1.5\%$).
The \emph{Cross-sessions Concat} model is more prone to overfitting when including these actions with weaker preference signals because it estimates weights for every action and takes the order of actions into account while e.g., the \emph{Cross-sessions Encode} model only accounts for the order of sessions.

Not surprisingly, the most negative impact (up to $-19.86\%$ in MAP) is obtained when actions with the object type ``item'' are removed.
Even if these are not the most frequent objects (see Tab.~\ref{tab:actions}), they are highly informative as they provide information on the user interests.
The most frequent objects are ``services'', twice more frequent than items, and their removal affects negatively performance even if with a less severe impact (up to $-6.61\%$ in MAP).

The second greatest negative impact is obtained when the actions from the section ``e-commerce'' are removed (up to $-16.77\%$ in MAP).
Again, this is not the most frequent section, but it is highly informative since the user needs to access the e-commerce section to inspect and compare different insurance products.
The most frequent section is the ``personal account'', $3$ times more frequent than the e-commerce section.
Its removal has a negative impact, but not as severe as for e-commerce (up to $-7\%$ in MAP).

In terms of action type, the removal of the type ``start'' has the greatest negative effect, even if limited with respect to the other two categories (up to $-4.77\%$ in MAP).
The types ``act'' and ``complete'' have a negligible impact (less than $1.5\%$), but they also represent a very sparse signal (less than $5\%$ of all types together). 

\section{Conclusions and Future Work}
\label{subsec:conclusions}

We have tackled the problem of recommendation of insurance products. This is a highly relevant industrial problem, to which no satisfying solution currently exists, because of the low number of items and sparsity of user interactions.
We propose three different cross-sessions recommendation models, which exploit both the current and past user sessions to predict items that the user will buy.
Our models take as input an ordered sequence of sessions each being a list of actions and model the dependencies across sessions using RNNs with GRU units.
Experimental results on a real world dataset show that our models outperform state-of-the-art baselines.
Further analysis confirms the positive effect of considering multiple past sessions and an ablation study shows that all considered action types are beneficial for the models.
Demographic features boost performance of the cross-sessions model, giving rise for future work on potential biases of demographic insurance recommendations.

As future work we further plan to run an $A/B$ testing experiment to evaluate our cross-sessions models with online users.
Furthermore, we will investigate the explainability of our models and we will generate user readable explanations to be shown when an item is recommended.

\begin{acks}
This paper is partially supported by the EU Horizon 2020 MSCA grant No. 893667.
\end{acks}

\bibliographystyle{ACM-Reference-Format}
\bibliography{sample}


\end{document}